\documentstyle[11pt,newpasp,twoside,epsf]{article}
\def\edcomment#1{\iffalse\marginpar{\raggedright\sl#1\/}\else\relax\fi}
\reversemarginpar

\newcommand{\mgii}{\mbox{${\rm Mg_{2}} $}}

        \newcommand{\alfa}{\mbox{$\alpha$-elements}}
        \newcommand{\alfe}{\mbox{$\alpha$-enhanced}}

	\newcommand{\Xsun}{\mbox{${\rm X_{\odot}}$}}
	\newcommand{\Zsun}{\mbox{${\rm Z_{\odot}}$}}

\begin{document}
\title{Indices for SSPs with enhanced mix of $\alpha$-elements}
 \author{Rosaria Tantalo}
\affil{Department of Astronomy, University of Padua,\\
Vicolo dell'Osservatorio 2, 35122 -- PADOVA Italy}

\begin{abstract}
In this poster line strength indices for new Single Stellar
Populations (SSP) based on the recent stellar models with chemical
compositions enhanced in the \alfa\ (Salasnich et al. 2000) are
presented. The SSP indices have been calculated taking into account
different methods to correct the {\it Standard-Worthey Indices} for
the real chemical composition of \alfe\ isochrones.
\end{abstract}

\section{Introduction}
Over the years, much effort has been spent to interpret main
properties of Ellipticals galaxies such as the age, metallicity, and
chemical properties. To this aim, Tantalo et al. (1998) (TCB98) have developed
the so-called $\Delta$-Method to get the metallicity, age, and {\it
enhancement-degree} for a sample of galaxies. Since the isochrones adopted by
TCB98 were with solar partition of elements, in this study by using isochrones
based on \alfe\ stellar models (see Salasnich et al. 2000 for all details on
these models) and calibrations of indices in which the effect of different
[$\alpha$/Fe] ratios are simultaneously taken into account, we seek to improve
upon a point of weakness in the previous analysis.

\section{How the different mixture of \alfa\ will affect the indices ?}

In presence of a certain amount of enhancement in the \alfa\
one has to suitably modify the relationship between the total
metallicity $Z$ and the iron content $[Fe/H]$. This can be expressed by
the following relation in which $\Gamma$ is the {\it Degree of Enhancement}

\begin{equation}
\left[ \frac{Fe}{H} \right] = \lg{\left(\frac{Z}{\Zsun}\right)} - \lg{\left(\frac{X}{\Xsun}\right)} + \Gamma
\label{feh}
\end{equation}

\noindent
For solar metallicity and solar-scaled isochrones ($\Gamma = 0.0$),
[Fe/H] is $=0.0$, whereas in the case of \alfe\ isochrones
with solar metallicity the real abundance is [Fe/H]$=-0.3557$.

In the analysis below, we have calculated two different sets
of SSPs indices following two different procedures for including the
effects of the enhancement of \alfa\ on the line strength indices. {\bf
BC}: by using the empirical calibration for \mgii\ given by Borges et al.
(1995) as in TCB98. {\bf TC}: by using the index corrections according eq.(6)
of Trager et al. (2000).

\section{Conclusions}

Fig.1 shows, the results for the \mgii\ index corrected according {\bf TC} and
{\bf BC} method in left-panel and right-panel respectively (see Tantalo et al.
2001, for more details). In both cases indices tend to flatten out at old ages
(say above 3Gyr), whereas they are strongly depend on metallicity and
enhancement.

\begin{figure}
\plottwo{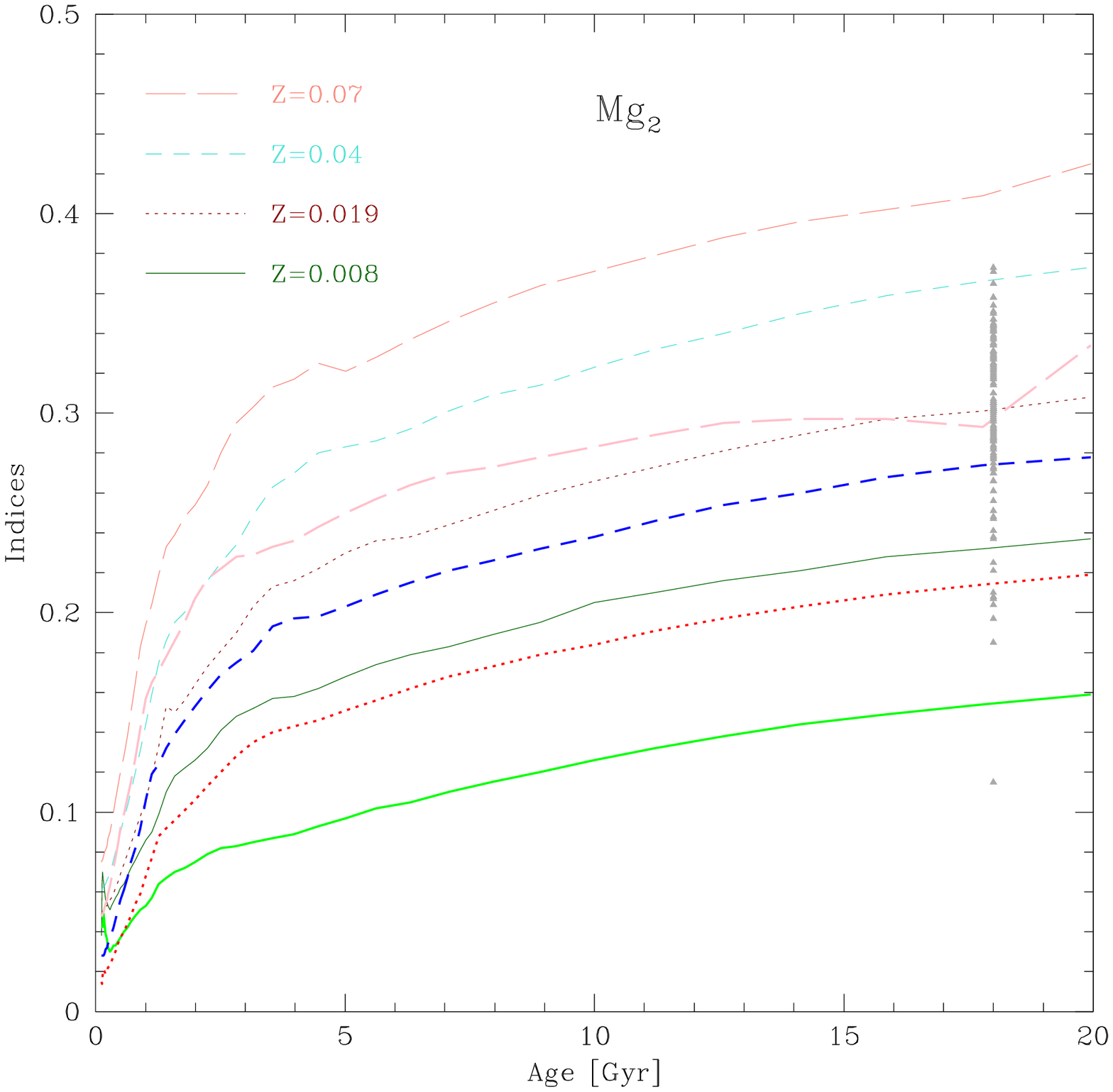}{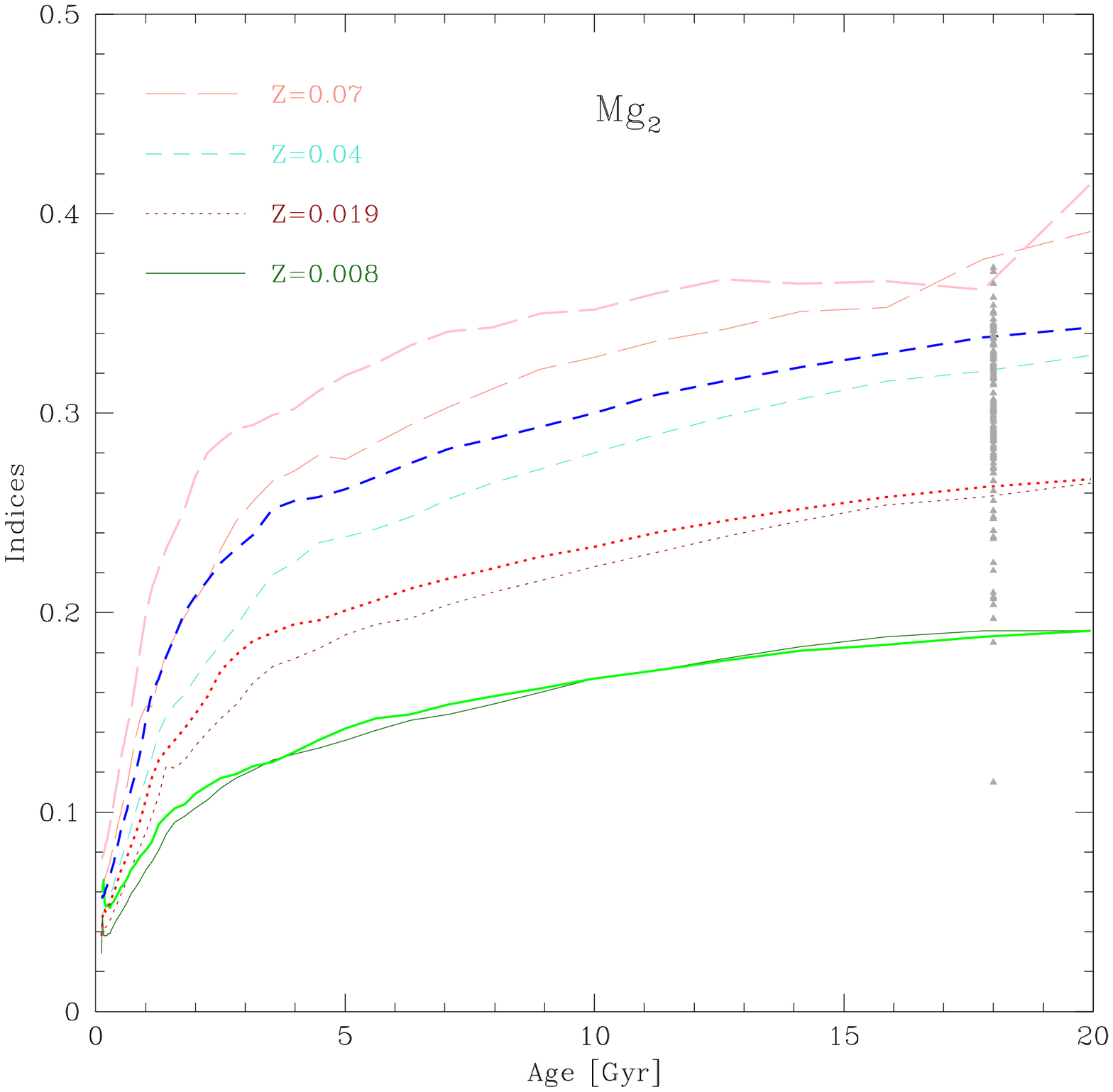}
\caption{Both panels shown the \mgii\ index as function of age for
solar-scaled and \alfe\ abundances, {\it thin} and {\it thick lines}
respectively calculated with {\bf TC} (left) and {\bf BC} (right) method.
{\it Solid}, {\it dotted}, {\it dashed} and {\it long-dashed} lines are for
Z=0.008, Z=0.019, Z=0.040 and Z=0.070 respectively. {\it Full-triangles}
visualizes the range of the Trager et al. (2000) data sample.}
\label{index}
\end{figure}

The major drawback of the new indices is their {\it age-degeneracy} beyond
3Gyr. Therefore removing the classical age-metallicity degeneracy is a
cumbersome affair. A preliminary determination of the age by means of
$\Delta$-Method of TCB98 does not lead either to a unique
solution or acceptable ages (too old).

Despite the above {\it age-degeneracy}, in Tantalo et al. (2001)
using a minimum-distance procedure, we have estimated the age, metallicity , and
$\Gamma$ for a sample of galaxies. The ranges spanned by age, metallicity, and
$\Gamma$ are now compatible with estimates derived from other, independent
methods (Trager et al. 2000).


\end{document}